\documentclass{article}

\usepackage{arxiv}

\usepackage{amsmath,amsfonts,amssymb}
\usepackage{array}
\usepackage{booktabs}
\usepackage{multirow}
\usepackage{tabularx}
\usepackage{graphicx}
\usepackage[caption=false,font=normalsize,labelfont=sf,textfont=sf]{subfig}
\usepackage{textcomp}
\usepackage{stfloats}
\usepackage{placeins}
\usepackage{url}
\usepackage{verbatim}
\usepackage{cite}
\usepackage{algorithm}
\usepackage{orcidlink}
\usepackage{algpseudocode}
\usepackage{pgfplots}
\usepackage{graphicx}
\usepackage{wrapfig}
\usepackage{pgfplotstable}
\usepackage{tikz}
\usepgfplotslibrary{groupplots}
\usetikzlibrary{positioning,arrows.meta,shapes.geometric,fit}
\pgfplotsset{compat=1.18}

\title{\textbf{RES-DARE}: Failure-Aware Expert Adaptation and Rollback-Safe Self-Repair for Intrusion Detection}

\author{
\textbf{Rahil Aftab}~\orcidlink{0009-0009-0983-4726}\\
Department of Computer Science\\
Jamia Hamdard\\
New Delhi 110062, India\\
\texttt{rahilaftab12@gmail.com}
\and
\textbf{Anyash Prasad}~\orcidlink{0009-0006-6684-5091}\\
Department of Computer Science\\
Kalinga Institute of Industrial Technology\\
Bhubaneswar, Odisha 751024, India\\
\texttt{anyashprasad@gmail.com}
\and
\textbf{Soumya Mazumdar}~\orcidlink{0009-0006-3521-9557}\\
Department of Computer Science and Business Systems\\
Gargi Memorial Institute of Technology\\
Affiliated to Maulana Abul Kalam Azad University of Technology\\
Balarampur, Mouza Beralia, Baruipur, Kolkata 700144, West Bengal, India\\
\texttt{reachme@soumyamazumdar.com}
\and
\textbf{Vineet Kumar Rakesh}~\orcidlink{0009-0000-7102-6564}\\
Engineering Science\\
Homi Bhabha National Institute\\
Anushaktinagar, Mumbai 400094, Maharashtra, India\\
Computer and Informatics Group\\
Variable Energy Cyclotron Centre\\
1/AF, Bidhannagar, Kolkata 700064, West Bengal, India\\
\texttt{vineet@vecc.gov.in}
\and
\textbf{Tapas Samanta}~\orcidlink{0000-0003-0521-0747}\\
Engineering Science\\
Homi Bhabha National Institute\\
Anushaktinagar, Mumbai 400094, Maharashtra, India\\
Computer and Informatics Group\\
Variable Energy Cyclotron Centre\\
1/AF, Bidhannagar, Kolkata 700064, West Bengal, India\\
\texttt{tsamanta@vecc.gov.in}
}

\begin{document}
\maketitle

\begin{abstract}
Intrusion detection systems are often trained under static benchmark conditions, although deployed network environments are affected by traffic drift, sensor noise, changing workloads, and evolving attack behaviour. Under such distribution shifts, static detectors may produce confident but incorrect predictions, leading to silent and unsafe failure modes. In this paper, RES-DARE (Recursive Evolving Specialists-Digital Adaptive Reasoning Engine) is proposed as a failure-aware continual intrusion detection framework with rollback-safe self-repair. Difficult, uncertain, and misclassified samples are treated as failure signals for expert specialisation rather than being discarded as noise. A supervised contrastive encoder, two-pass expert router, failure-buffer mechanism, HDBSCAN-based failure-region discovery, and trust-risk monitor are integrated to support adaptive IDS behaviour. AEHM-v2 is introduced as a rollback-safe repair mechanism, where candidate adaptations are provisionally activated and committed only when macro-F1 is preserved or improved while trust risk remains stable. Otherwise, the system is rolled back to its last validated state. RES-DARE is evaluated on CICIDS2017, UNSW-NB15, and TON\_IoT, achieving macro-F1 scores of 0.9850, 0.9736, and 0.9691, respectively. Under Gaussian feature corruption at strength 0.10, RES-DARE retains an Attack-F1 of 0.7920 on CICIDS2017 and achieves near-zero catastrophic forgetting with F = 0.0015. The results show that RES-DARE improves robustness, warning capability, and deployment safety under degraded conditions.
\end{abstract}


\section{Introduction}

Modern intrusion detection systems are increasingly deployed in environments where traffic statistics drift, sensors degrade, and attacker behaviour changes faster than static training assumptions. In this setting, a model that scores well on a clean benchmark may still fail dangerously: the most concerning case is not merely low accuracy, but confident failure under distribution shift.

This paper presents RES-DARE (Recursive Evolving Specialists -- Digital Adaptive Reasoning Engine), a failure-aware continual IDS framework designed around three principles: failures should be treated as signals for specialisation rather than discarded as noise; adaptation must be validated before it is committed; and evaluation must measure operational robustness alongside clean metrics. 

RES-DARE combines a frozen representation encoder, dynamic expert routing, HDBSCAN-based failure-region discovery to identify clusters of difficult samples, trust-risk monitoring, and Adaptive Expert Health Management version 2 (AEHM-v2). AEHM-v2 acts as a rollback-safe self-repair controller: when degradation signals accumulate, a repair adapter is activated provisionally, validated, and retained only if it improves or preserves macro-F1 while reducing or preserving trust risk.

The main contributions of this paper are:
\begin{itemize}
\renewcommand{\labelitemi}{$\bullet$}
\item A failure-aware continual intrusion detection framework, named RES-DARE, that treats difficult, uncertain, and misclassified traffic samples as informative signals rather than noise.
\item A dynamic expert-specialisation mechanism leveraging contrastive representations, two-pass routing, failure-buffer analysis, and HDBSCAN to allocate model capacity toward persistently difficult traffic regions.
\item A rollback-safe self-repair mechanism, AEHM-v2, ensuring that model adaptation is provisionally evaluated and committed only when macro-F1 is preserved or improved and trust risk remains stable.
\item A trust-risk monitoring strategy incorporating routing entropy, confidence behaviour, failure density, prototype drift, and expert activity to detect operational degradation.
\item Systematic evaluation across three public datasets (CICIDS2017, UNSW-NB15, TON\_IoT) capturing clean performance, robustness, silent failure rate, warning capability, and catastrophic forgetting.
\end{itemize}

\section{Related Work}

Network intrusion detection has matured considerably on clean benchmarks, with tree-based methods such as Random Forest and XGBoost remaining strong baselines on engineered flow features \cite{breiman2001random,chen2016xgboost}. Surveys confirm that static models perform well under controlled conditions but break down under concept drift, class imbalance, and distribution shift \cite{Gamage2020,Yang2022SLR,Shyaa2024,Zhong2024,Komarchesqui2026DriftSurvey}. Once deployed, a static model requires manual retraining when traffic shifts and gives no direct warning signal when its internal predictions become unreliable.

Representation and contrastive learning methods improve generalisation by organising embeddings around class boundaries \cite{khosla2020supervised,LopezMartin2022,Li2024Contrastive}. mixture-of-experts (MoE) architectures extend this idea by allocating capacity to different input regions \cite{shazeer2017outrageously,Mu2025MoESurvey,Gulmez2026DynaMoE,Luo2026Sphere,Vivek2025HybridXAI}. However, specialisation alone does not solve stale representations or unsafe adaptation: the system still needs health monitoring and a way to reject harmful updates.

Robustness and out-of-distribution work measures degradation under shift \cite{Corsini2023CyberOOD,Wang2024FeatureShift,Wang2023RobustnessNIDS,Farrukh2024AISNIDS}. Continual learning extends operation through replay, regularisation, pruning, and architectural growth \cite{parisi2019continual,Golkar2019Pruning,Douillard2022DyTox,Li2025OpenWorldCL,Karn2021LwF,Zhang2024EvoFedIDS,Niu2026RepShield,Fuhrman2026ACORN,Banerjee2026Forgetting,Guo2025ContinualIDS,Soltani2024MultiAgent}. Uncertainty or anomaly-monitoring methods expose unreliable predictions \cite{Hwang2020Unsupervised,Naeimi2025REMA,Talpini2024,Medjadba2025Bayesian,Fernando2022DWBL,gal2016dropout}. RES-DARE addresses the gap between these threads by using failures to drive expert specialisation, trust monitoring to guide repair decisions, and AEHM-v2 to validate adaptation before it is committed or rolled back.

\section{Methodology}

\subsection{System Overview}

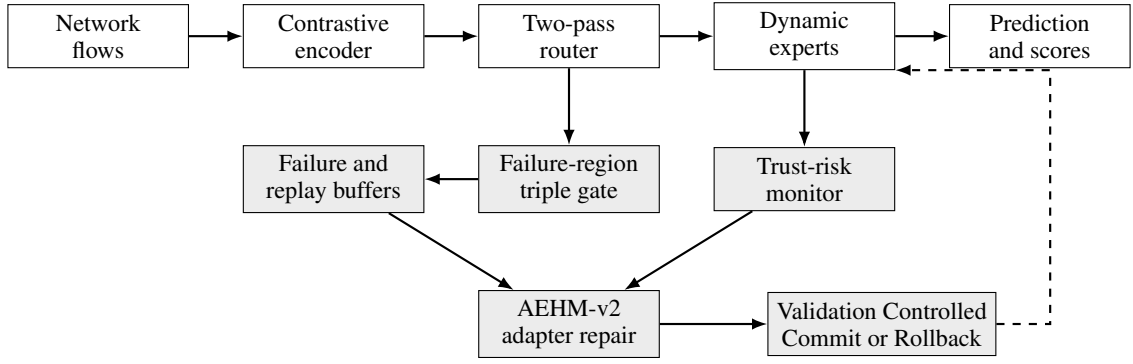
\begin{figure}[t]
\centering
\resizebox{0.90\textwidth}{!}{%
\begin{tikzpicture}[
    node distance=5mm and 7mm,
    box/.style={draw, rectangle, minimum width=2.35cm, minimum height=0.82cm, align=center, fill=white, font=\small},
    graybox/.style={box, fill=gray!15},
    arrow/.style={-{Latex[length=2mm]}, thick},
    safe/.style={-{Latex[length=2mm]}, thick, dashed}
]
\node[box] (flow) {Network\\flows};
\node[box, right=of flow] (enc) {Contrastive\\encoder};
\node[box, right=of enc] (router) {Two-pass\\router};
\node[box, right=of router] (expert) {Dynamic\\experts};
\node[box, right=of expert] (pred) {Prediction\\and scores};

\node[graybox, below=10mm of enc] (buffer) {Failure and\\replay buffers};
\node[graybox, below=10mm of router] (gate) {Failure-region\\triple gate};
\node[graybox, below=10mm of expert] (trust) {Trust-risk\\monitor};
\node[graybox, below=10mm of gate] (repair) {AEHM-v2\\adapter repair};
\node[graybox, right=14mm of repair] (val) {Validation Controlled\\Commit or Rollback};

\draw[arrow] (flow) -- (enc);
\draw[arrow] (enc) -- (router);
\draw[arrow] (router) -- (expert);
\draw[arrow] (expert) -- (pred);
\draw[arrow] (router) -- (gate);
\draw[arrow] (gate) -- (buffer);
\draw[arrow] (buffer) -- (repair);
\draw[arrow] (expert) -- (trust);
\draw[arrow] (trust) -- (repair);
\draw[arrow] (repair) -- (val);
\draw[safe] (val.east) -- ++(7mm,0) |- (expert.south east);
\end{tikzpicture}%
}
\caption{RES-DARE architecture overview. The primary inference pipeline feeds telemetry to the trust monitor and failure gates, triggering provisional AEHM repairs that validate performance before committing updates.}
\label{fig:architecture}
\end{figure}

The RES-DARE pipeline begins with a supervised contrastive encoder that maps tabular flow features into an embedding space, as shown in Fig.~\ref{fig:architecture}. A two-pass router assigns embeddings to expert heads. Failure scoring collects hard or uncertain samples into a bounded buffer. Persistent failure regions can spawn new experts, and trust-risk telemetry monitors whether routing behaviour, confidence, and failure density indicate degradation.

\begin{figure}[t]
\centering
\resizebox{0.92\textwidth}{!}{%
\begin{tikzpicture}[
    node distance=4mm and 6mm,
    box/.style={draw, rectangle, minimum width=2.15cm, minimum height=0.72cm, align=center, fill=white, font=\scriptsize},
    graybox/.style={box, fill=gray!15},
    arrow/.style={-{Latex[length=1.8mm]}, thick},
    safe/.style={-{Latex[length=1.8mm]}, thick, dashed}
]
\node[box] (features) {Preprocessed\\features};
\node[box, right=of features] (mlp) {MLP encoder\\256-256-128};
\node[box, right=of mlp] (emb) {64-d\\embedding};
\node[box, right=of emb] (routeflow) {Router\\scores};
\node[box, right=of routeflow] (heads) {Expert MLP\\heads};
\node[box, right=of heads] (scores) {Class\\scores};

\node[graybox, below=9mm of emb] (fail) {Failure\\embeddings};
\node[graybox, right=of fail] (hdb) {HDBSCAN\\region gate};
\node[graybox, below=9mm of heads] (telemetry) {Loss, confidence\\routing entropy};
\node[graybox, below=8mm of hdb] (growth) {Expert\\growth/update};
\node[graybox, below=8mm of telemetry] (adapter) {AEHM-v2\\adapter};
\node[graybox, right=of adapter] (repairval) {Repair-validation\\batch};

\draw[arrow] (features) -- (mlp);
\draw[arrow] (mlp) -- (emb);
\draw[arrow] (emb) -- (routeflow);
\draw[arrow] (routeflow) -- (heads);
\draw[arrow] (heads) -- (scores);
\draw[arrow] (emb) -- (fail);
\draw[arrow] (fail) -- (hdb);
\draw[arrow] (hdb) -- (growth);
\draw[arrow] (heads) -- (telemetry);
\draw[arrow] (telemetry) -- (adapter);
\draw[arrow] (growth) -- (adapter);
\draw[arrow] (adapter) -- (repairval);

% --- PERFECTLY ALIGNED, NON-OVERLAPPING ROUTING LOOPS ---
\coordinate (gapY) at ([yshift=4.5mm]telemetry.north);

\coordinate (leftchannel) at ([xshift=4.5mm]growth.east);
\draw[arrow] (growth.east) -- (leftchannel |- growth.east) 
                           -- (leftchannel |- gapY) 
                           -- ([xshift=-6mm]heads.south |- gapY) 
                           -- ([xshift=-6mm]heads.south);

\coordinate (rightchannel) at ([xshift=6mm]repairval.east);
\draw[safe] (repairval.east) -- (rightchannel |- repairval.east) 
                             -- (rightchannel |- gapY) 
                             -- ([xshift=6mm]heads.south |- gapY) 
                             -- ([xshift=6mm]heads.south);
\end{tikzpicture}%
}
\caption{RES-DARE layer flow. Preprocessed flow features pass through the MLP encoder and expert heads; HDBSCAN and AEHM operate on failure evidence and validation feedback rather than replacing the main classifier.}
\label{fig:layerflow}
\end{figure}
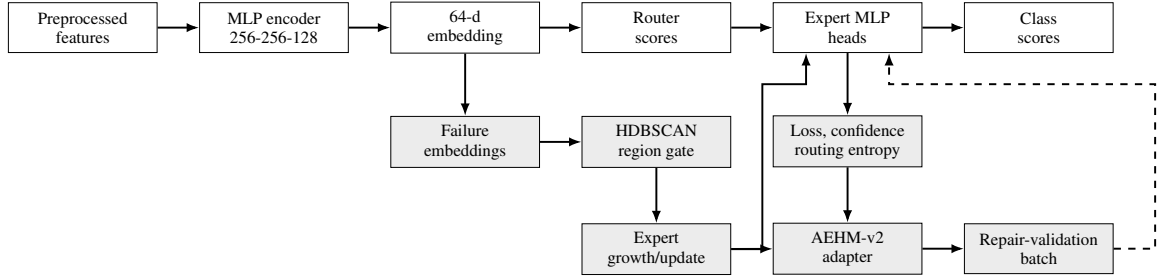

The layer-level data flow is shown in Fig.~\ref{fig:layerflow}: preprocessed features enter the MLP encoder, embeddings feed the router and expert heads, HDBSCAN analyses failure embeddings, and AEHM watches trust signals before committing or rolling back an adapter.

\subsection{Encoder and Failure Scoring}

The encoder is a MLP (multilayer perceptron) trained with supervised contrastive loss, which forces embeddings of similar traffic types closer together while pushing distinct attacks further apart, and cross-entropy. After warmup, the embedding space is used by expert routing and failure-region discovery. For a sample $x_i$, the model produces embedding $z_i$ and class probabilities $p_i$. Equation~\ref{eq:failure_score} defines the failure score used to combine prediction confidence and distance from assigned expert prototypes in equation~\ref{eq:failure_score}.
\begin{equation}
s_i = w_c (1 - \max p_i) + w_d d_M(z_i, \mu_e),
\label{eq:failure_score}
\end{equation}
where $w_c$ and $w_d$ are fixed validation-tuned weights; both are set to 1.0 in the reported runs. The term $d_M$ is a Mahalanobis-style distance to expert prototype $\mu_e$, a metric that scales distance by the variance of the cluster rather than treating all directions equally. During cold start, the implementation falls back to normalised Euclidean distance until enough samples are available for stable covariance estimates.

\subsection{Expert Routing and Growth}

Each expert is a compact classifier head over encoder embeddings. Routing is two-pass: a fast prototype-similarity pass selects a candidate expert, while a second pass uses uncertainty and prototype distance to decide whether the sample should be handled directly or added to the failure buffer. New experts are created only when a failure region passes the triple gate in equation~\ref{eq:triple_gate}.
\begin{equation}
G = G_{\mathrm{density}} \wedge G_{\mathrm{persistence}} \wedge G_{\mathrm{novelty}} .
\label{eq:triple_gate}
\end{equation}
The default implementation uses HDBSCAN \cite{campello2013density} to identify persistent clusters of failed samples -- regions of the input space where the model consistently struggles. An alternative trigger based on maximum mean discrepancy (MMD), a statistical test for detecting whether two sets of samples come from different distributions, is available for ablation experiments; memory-efficient approximations are used to keep it computationally tractable \cite{gretton2012kernel}.

\subsection{Trust Monitoring and AEHM Repair}

The trust monitor tracks signals such as routing entropy, prototype drift, uncertainty behaviour, expert disagreement, and failure density. Routing entropy measures uncertainty in expert assignment and serves as a primary trigger for degradation alerts within the health management loop. Higher routing entropy indicates that multiple experts appear similarly suitable for a sample, suggesting increasing ambiguity or distributional shift. Equation~\ref{eq:trust_risk} defines the trust-risk score used to combine these telemetry streams equation~\ref{eq:trust_risk}.
\begin{equation}
R_t =
\mathrm{clip}\left(
\omega_L L_t +
\omega_E E_t +
\omega_F F_t +
\omega_P P_t +
\omega_A A_t +
\omega_S S_t,
0,1
\right),
\label{eq:trust_risk}
\end{equation}
where $L_t$ denotes the normalized loss term, $E_t$ denotes routed error or uncertainty behaviour, $F_t$ denotes the failure-buffer ratio, $P_t$ denotes the second-pass routing ratio, $A_t$ denotes active-expert saturation, and $S_t$ denotes expert-spawn or expert-health activity. The weights $\omega$ are fixed implementation parameters. Lower $R_t$ indicates a healthier deployment state.

AEHM validation-controlled repair was evaluated across CICIDS2017, UNSW-NB15, and TON\_IoT\_SMALL\_5P. The strict repair validation logic acts as a commit/rollback controller: when degradation signals accumulate, a provisional repair adapter is activated and evaluated. The update is formally committed to the system state only if it strictly improves performance metrics without increasing deployment risk.

The AEHM lifecycle has five steps: monitor health, accumulate failure evidence, train a repair adapter, deploy it provisionally, and validate performance and trust risk. The live model is not touched until validation completes. If the repaired model improves or preserves macro-F1 while holding trust risk stable, the adapter is committed and the system resumes normal operation. If either criterion fails, the adapter is discarded and the system rolls back to its last validated state. Crucially, the cycle then returns to monitoring rather than retrying indefinitely, preventing repair loops from compounding a degraded state. This lifecycle is summarized in Fig.~\ref{fig:architecture}.

The commit decision is formalized after provisional deployment. Let $M_{\mathrm{old}}$ and $R_{\mathrm{old}}$ denote the macro-F1 and trust-risk before candidate activation, and let $M_{\mathrm{new}}$ and $R_{\mathrm{new}}$ denote the corresponding values after candidate validation. The strict validation rule used in the reported AEHM-v2 repair traces is defined as mentioned in equation~\ref{eq:commit_rule}.
\begin{equation}
    \Delta M = M_{\mathrm{new}} - M_{\mathrm{old}}, 
    \qquad
    \Delta R = R_{\mathrm{new}} - R_{\mathrm{old}},
\end{equation}
\begin{equation}
    \mathrm{Commit} =
    \begin{cases}
    1, & \Delta M > 0 \ \mathrm{and}\ \Delta R \leq 0,\\
    0, & \mathrm{otherwise}.
    \end{cases}
    \label{eq:commit_rule}
\end{equation}
The encoder layer sizes, optimizer, batch size, routing thresholds, and repair settings used in the reported revision runs are summarized in Table~\ref{tab:settings}.

\begin{table}
\centering
\caption{Key implementation settings used in the reported RES-DARE revision runs.}
\label{tab:settings}
\begin{tabular}{|l|p{8.2cm}|}
\hline
Component & Setting \\
\hline
Encoder & MLP encoder with hidden dimensions [256, 256, 128] and 64-dimensional embedding. \\
Training & Adam optimizer for encoder training with learning rate $1\times10^{-3}$; supervised contrastive and classification objectives; batch size 256 in the reported revision runs; maximum 8 epochs for the repeated-seed stability experiments. \\
Phase-3 routing & Two-pass expert routing enabled; trust monitoring enabled; LSTM-based trust option enabled in the revision runs. \\
MoE training & MoE training maximum samples 4096, MoE batch size 128, specialist training budget 3 epochs. \\
Expert growth & HDBSCAN-based spawning, minimum cluster size 50, persistence threshold 0.65, novelty threshold 0.70, maximum active experts 6 in the repeated-seed revision runs. \\
AEHM buffer and repair & Bounded AEHM buffer of 8192 samples; candidate adapter training uses logged high-confidence assignments and available labelled/assigned samples; commit or rollback is decided by validation macro-F1 and trust-risk. \\
TON\_IoT repeated-seed subset & TON\_IoT\_SMALL\_5P, a stratified 5\% subset containing 206,401 samples. \\
\hline
\end{tabular}
\end{table}

\section{Experimental Setup}

In this experiment, we used three public intrusion-detection datasets: CICIDS2017 \cite{sharafaldin2018cicids}, UNSW-NB15 \cite{moustafa2015unsw}, and TON\_IoT, also referred to as ToN-IoT in the dataset literature \cite{moustafa2021ton}. CICIDS2017 represents enterprise-style benign and attack traffic, UNSW-NB15 contains modern attack categories mixed with normal traffic, and TON\_IoT represents IoT/edge network conditions. Experiments were run on an enterprise computing cluster equipped with dual NVIDIA L40S GPUs (Ada Lovelace architecture), each configuring 18,176 CUDA cores, 142 third-generation RT Cores, and 48\,GB of GDDR6 ECC tensor memory, operating under a PCIe Gen 4.0 $\times$16 interface with NVIDIA CUDA 12.6 drivers. For TON\_IoT, final clean detection and binary attack-detection metrics were computed on the full evaluation set of 4,128,039 samples. For computationally intensive robustness, uncertainty, feature-occlusion, and ablation studies, we used a stratified 5\% subset, denoted TON\_IoT\_SMALL\_5P, containing 206,401 samples.

All datasets use the same preprocessing pipeline for RES-DARE and the baselines. Categorical fields are encoded when present, numeric features are normalized using training-set statistics, and the same train, validation, and test partitions are kept fixed across methods. Static baselines include Logistic Regression (LogReg) \cite{cox1958regression}, Linear SVM (Support Vector Machine) \cite{cortes1995support}, Isolation Forest (IF) \cite{liu2008isolation}, Random Forest (RF) \cite{breiman2001random}, MLP (multilayer perceptron) \cite{rumelhart1986learning}, and XGBoost (eXtreme Gradient Boosting) \cite{chen2016xgboost}. All static baselines use the same preprocessed splits as RES-DARE. Logistic Regression and MLP use standard scaling; Random Forest uses 100 trees with balanced subsampling; MLP uses hidden layers (128,64), batch size 512, early stopping, and learning rate $10^{-3}$; XGBoost uses 200 trees, depth 6, learning rate 0.08, and histogram tree construction. Confidence values for silent-failure analysis are taken from native probability outputs where available; for models without calibrated probabilities, the reported confidence should be interpreted as a model-internal warning signal rather than a calibrated cross-model probability.

Macro-F1 is the unweighted mean of per-class F1 scores, and Binary Attack-F1 collapses all attack classes into a single positive label. AUROC measures ranking quality independent of a fixed threshold. Corruption robustness reports Attack-F1 under synthetic perturbations, silent failure rate measures the fraction of wrong predictions made with confidence at least 0.90, and warning AUROC measures how well uncertainty signals separate correct from incorrect predictions. Equation~\ref{eq:forgetting} computes forgetting as the average drop from the best historical attack/minority recall to the final recall after later adaptation:
\begin{equation}
\mathcal{F} = \frac{1}{T-1}\sum_{i=1}^{T-1}\left(\max_{t<T} R_i(t) - R_i(T)\right),
\label{eq:forgetting}
\end{equation}
where $R_i(t)$ is attack/minority recall on segment $i$ after step $t$.

Labels are assumed to become available periodically during the repair phase, for example through a human-verified batch; continuous real-time labelling is not required. The system does not assume prior knowledge of attack types that are entirely absent from training, so novel attacks are flagged as uncertain rather than pre-classified. Corruption experiments are applied after preprocessing to the normalized feature matrix, where categorical fields have already been encoded. Gaussian corruption adds zero-mean noise to all feature columns with standard deviation equal to the selected strength times each feature's empirical standard deviation; values are not clipped after corruption. The 0.05 and 0.10 strengths are used as mild and moderate normalized-noise settings. Mask corruption sets randomly selected feature entries to zero, and scale corruption multiplies randomly selected feature columns by factors sampled uniformly from 0.5 to 1.5.

\subsection{Continual Learning Experimental Protocol}
To evaluate RES-DARE under deployment-like conditions, the Phase-3 evaluation is formulated as a sequential validation procedure over steps $t=1,2,\ldots,T$. Each step represents a new evaluation pass or stream segment processed by the currently active RES-DARE state. The protocol separates prediction-time telemetry, delayed label use, candidate adaptation, and validation-controlled repair.

\textbf{Stream construction:} The processed evaluation records are consumed in a deterministic sequential order. CICIDS2017 and UNSW-NB15 are evaluated using their full processed evaluation streams. For TON\_IoT, the final clean detection result is reported on the full evaluation set, while repeated-seed and computationally expensive revision experiments use TON\_IoT\_SMALL\_5P, a stratified 5\% subset containing 206,401 samples.

\textbf{Prediction and telemetry collection:} At step $t$, the active model processes the incoming samples without using class labels for routing or prediction. For each sample, RES-DARE records the predicted class, confidence, uncertainty-related quantities, routing entropy, expert assignment, second-pass routing status, and failure/health indicators. These signals are used to update the live AEHM state and the bounded adaptation buffer. High-confidence model assignments may be retained for stabilization; however, they are not treated as verified correct predictions until labels are available.

\textbf{Delayed label use:} Labels are used only after the prediction pass for metric computation, validation, and controlled maintenance. This simulates a practical intrusion-detection setting in which labels may become available after analyst verification, delayed incident triage, or batch SIEM processing. Thus, labels are not used to make the initial online prediction, but they are used to evaluate whether a candidate repair should be accepted.

\textbf{Expert-routing and candidate detection:} Failure-prone samples are identified from the logged health signals, including confidence, uncertainty, routing behaviour, second-pass failures, and clustering diagnostics. In the reported revision runs, expert creation was controlled using HDBSCAN-based spawning with minimum cluster size 50, persistence threshold 0.65, novelty threshold 0.70, and a maximum active-expert cap of 6. These thresholds govern candidate expert spawning and should not be interpreted as a single universal failure-score threshold.

\textbf{Candidate adaptation:} When the AEHM health gate is activated, a provisional adapter is trained using samples accumulated in the bounded AEHM buffer together with stabilizing replay information. The implementation logs the buffer size, high-confidence sample count, number of assignment classes, number of training samples used, training loss, and adapter-training status. In the reported runs, the adapter-training routine used a small fixed training budget of three epochs.

\textbf{Validation-controlled commit or rollback:} A candidate adapter is not blindly deployed after training. Instead, it is evaluated against the pre-candidate state using validation macro-F1 and trust-risk. Let $M_{\mathrm{old}}$ and $R_{\mathrm{old}}$ denote the macro-F1 and trust-risk before candidate activation, and let $M_{\mathrm{new}}$ and $R_{\mathrm{new}}$ denote the corresponding values after candidate validation. The strict validation rule used in the reported AEHM-v2 repair traces is
\begin{equation}
    \Delta M = M_{\mathrm{new}} - M_{\mathrm{old}}, 
    \qquad
    \Delta R = R_{\mathrm{new}} - R_{\mathrm{old}},
\end{equation}
\begin{equation}
    \mathrm{Commit} =
    \begin{cases}
    1, & \Delta M > 0 \ \mathrm{and}\ \Delta R \leq 0,\\
    0, & \mathrm{otherwise}.
    \end{cases}
    \label{eq:protocol_commit_rule}
\end{equation}
If the rule is satisfied, the candidate state is committed. Otherwise, the candidate adapter is rejected and the system rolls back to the pre-candidate state. This is the core AEHM safety mechanism: candidate repairs are validated before deployment rather than blindly accepted.

\textbf{Forgetting assessment:} Continual-learning stability is measured using minority/attack recall forgetting. For each earlier phase $i$, the best historical recall is compared with the recall after later adaptation. The forgetting score is computed as
\begin{equation}
    \mathcal{F}
    =
    \frac{1}{T-1}
    \sum_{i=1}^{T-1}
    \left(
    \max_{t<T} R_i(t) - R_i(T)
    \right),
    \label{eq:protocol_forgetting}
\end{equation}
where $R_i(t)$ is the minority/attack recall for phase $i$ after step $t$. Lower $\mathcal{F}$ indicates better retention of previously learned attack behaviour.

\section{Results}

\subsection{Clean Detection}

The first question for any IDS is simple: can it correctly identify attacks under normal, clean conditions? Table~\ref{tab:clean} reports macro-F1, Attack-F1, binary macro-F1, and AUROC for RES-DARE across all three datasets. The system achieves macro-F1 above 0.96 on every dataset, meaning it correctly identifies the vast majority of attack types including minority classes that simpler models tend to miss. In Table~\ref{tab:static_clean}, RES-DARE is compared against static baselines. XGBoost and Random Forest score higher on clean macro-F1 -- this is expected, as tree-based models are well suited to tabular flow features under stable conditions. RES-DARE is not designed to win on clean accuracy alone. What it trades in peak clean performance it recovers in operational robustness.

To check run-to-run stability, Phase-3 evaluation was repeated across three random seeds. Best-checkpoint Macro-F1 was $0.9713 \pm 0.0044$ on CICIDS2017, $0.9615 \pm 0.0010$ on UNSW-NB15, and $0.9505 \pm 0.0028$ on the TON\_IoT 5\% subset. Best-checkpoint AUROC was also stable at $0.9981 \pm 0.0006$, $0.9951 \pm 0.0002$, and $0.9907 \pm 0.0002$, respectively. These repeated-seed results provide empirical stability evidence and do not replace the sealed main checkpoints reported in Table~\ref{tab:clean}. The diagnostic execution logs and initialization scripts are provided alongside the supplementary material artifacts.

\begin{table}[htbp]
\begin{minipage}{0.48\textwidth}
\centering
\caption{Clean performance of final RES-DARE checkpoints ($\uparrow$).}
\label{tab:clean}
\resizebox{\textwidth}{!}{%
\begin{tabular}{|l|c|c|c|c|}
\hline
Dataset & Macro-F1 & Attack-F1 & Bin Macro & AUROC \\
\hline
CICIDS17 & 0.9850 & 0.9752 & 0.9846 & 0.9993 \\
UNSW15   & 0.9736 & 0.9804 & 0.9730 & 0.9972 \\
TON\_IoT & 0.9691 & 0.9611 & 0.9684 & 0.9961 \\
\hline
\end{tabular}%
}
\end{minipage}
\hfill
\begin{minipage}{0.48\textwidth}
\centering
\caption{Clean macro-F1 comparison with static baselines ($\uparrow$).}
\label{tab:static_clean}
\resizebox{\textwidth}{!}{%
\begin{tabular}{|l|c|c|c|}
\hline
Method & CICIDS17 & UNSW15 & TON\_IoT \\
\hline
XGBoost       & \textbf{0.9987} & \textbf{0.9867} & \textbf{0.9787} \\
RF            & 0.9979          & 0.9813          & 0.9778 \\
MLP           & 0.9710          & 0.9699          & 0.9745 \\
RES-DARE      & 0.9850          & 0.9736          & 0.9691 \\
Linear SVM    & 0.8755          & 0.8813          & 0.9033 \\
LogReg        & 0.7709          & 0.8743          & 0.8839 \\
Isol Forest   & 0.5777          & 0.2913          & 0.4708 \\
\hline
\end{tabular}%
}
\end{minipage}
\end{table}

\subsection{Comparison with QCL-IDS}

QCL-IDS is a recent continual IDS method that incorporates quantum-inspired components to stabilise learning over time \cite{zhu2026qclids}. Because QCL-IDS was not rerun under our preprocessing and split protocol, Table~\ref{tab:qcl} should be read as an approximate reference comparison against reported Attack-F1 values rather than a controlled same-pipeline reproduction. Under this limitation, RES-DARE is higher by 3.12 percentage points on CICIDS2017 and 3.94 points on UNSW-NB15.

\subsection{Rollback-Safe Self-Repair}
A core claim of RES-DARE is that adaptive repair should not be blindly applied. Table~\ref{tab:aehm} reports validation-controlled repair behaviour across CICIDS2017, UNSW-NB15, and the stratified TON\_IoT subset. On CICIDS2017, the strict AEHM-v2 repair trace produced a beneficial commit: macro-F1 improved and trust-risk decreased. On UNSW-NB15, validation showed that macro-F1 decreased from 0.9621 to 0.9528 and trust-risk increased, so the controller rejected the candidate adapter. On the TON\_IoT subset, trust-risk decreased but macro-F1 still fell from 0.9503 to 0.9447; because the performance criterion was not satisfied, the candidate was again rolled back. These results support the safety performance of the validation loop: beneficial repairs are securely committed, while harmful updates are prevented from destabilizing the active model deployment.

\begin{table}[htbp]
\begin{minipage}{0.45\textwidth}
\centering
\caption{Binary Attack-F1 comparison against QCL-IDS reference ($\uparrow$).}
\label{tab:qcl}
\resizebox{\textwidth}{!}{%
\begin{tabular}{|l|c|c|c|}
\hline
Dataset    & QCL-IDS & RES-DARE & Diff     \\
\hline
CICIDS17 & 0.9440  & \textbf{0.9752} & \textbf{+0.0312} \\
UNSW15   & 0.9410  & \textbf{0.9804} & \textbf{+0.0394} \\
\hline
\end{tabular}%
}
\end{minipage}
\hfill
\begin{minipage}{0.52\textwidth}
\centering
\caption{AEHM validation repair traces (TON\_IoT uses 5\% subset).}
\label{tab:aehm}
\resizebox{\textwidth}{!}{%
\begin{tabular}{|l|c|c|c|c|c|}
\hline
Dataset & Pre-F1 & Post-F1 & $\Delta$ F1 & $\Delta$ risk & Decision \\
\hline
CICIDS17 & 0.9605 & 0.9850 & \textbf{+0.0246} & \textbf{-0.0146} & Commit \\
UNSW15   & 0.9621 & 0.9528 & -0.0093 & +0.0073 & Rollback \\
TON\_IoT & 0.9503 & 0.9447 & -0.0056 & \textbf{-0.0176} & Rollback \\
\hline
\end{tabular}%
}
\end{minipage}
\end{table}

\subsection{Ablation Studies}

\subsubsection{MoE Routing Ablation.}
To isolate its contribution, we evaluate a variant where all samples are handled by a single classifier head over the same encoder embeddings, with no routing or specialisation. As shown in Table~\ref{tab:moe}, removing routing reduces macro-F1 across all three datasets. The largest drop occurs on TON\_IoT, where the gain from full RES-DARE is 0.0165 macro-F1 points. This is consistent with TON\_IoT containing more heterogeneous traffic patterns, where dynamic specialisation has the most to offer.

\subsubsection{Forgetting Ablation.}
Catastrophic forgetting is a practical concern for any system that adapts over time. Table~\ref{tab:forgetting} reports forgetting $\mathcal{F}$ across four variants of RES-DARE with components progressively removed. The full system achieves near-zero forgetting at $\mathcal{F}=0.0015$. Load balancing is the single largest contributor -- removing it raises forgetting to 0.1225, an 80-fold increase. Removing the uncertainty gate or replay/distillation each raises forgetting into the 0.05 range.

\begin{table}[htbp]
\begin{minipage}{0.48\textwidth}
\centering
\caption{No-MoE ablation ($\uparrow$).}
\label{tab:moe}
\resizebox{\textwidth}{!}{%
\begin{tabular}{|l|c|c|c|}
\hline
Dataset    & RES-DARE & No-MoE & Gain \\
\hline
CICIDS17 & 0.9850 & 0.9724 & \textbf{+0.0127} \\
UNSW15   & 0.9736 & 0.9657 & \textbf{+0.0080} \\
TON\_IoT & 0.9691 & 0.9526 & \textbf{+0.0165} \\
\hline
\end{tabular}%
}
\end{minipage}
\hfill
\begin{minipage}{0.48\textwidth}
\centering
\caption{Forgetting ablation ($\downarrow$).}
\label{tab:forgetting}
\resizebox{\textwidth}{!}{%
\begin{tabular}{|l|c|}
\hline
Variant                  & Forgetting $\mathcal{F}$ \\
\hline
Full RES-DARE            & \textbf{0.0015} \\
No uncertainty gate      & 0.0568 \\
No replay/distillation   & 0.0538 \\
No load balancing        & 0.1225 \\
\hline
\end{tabular}%
}
\end{minipage}
\end{table}

\subsection{Robustness Under Corruption}

Real deployments are rarely clean. Sensors degrade, network conditions introduce noise, and feature extraction pipelines can produce malformed values. To simulate this, we apply synthetic Gaussian noise to input features at two strengths and measure operational robustness parameters. Table~\ref{tab:gaussian} compiles both the remaining binary Attack-F1 metrics and the resulting silent failure rates under progressive distortion to provide a unified overview of system degradation. 

Static baselines collapse sharply under noise -- XGBoost drops to 0.3410 on CICIDS2017 at strength 0.10, barely above random. RES-DARE retains 0.7920 under the same conditions, a gap of 0.4510. This shows that RES-DARE continues to function meaningfully when operational conditions deteriorate. Furthermore, while models like XGBoost and Logistic Regression accumulate silent failures rapidly under noise, as shown in the silent failure columns (SF) of Table~\ref{tab:gaussian}, RES-DARE keeps silent failure rates consistently low across all datasets and corruption strengths.

\begin{table}[htbp]
\centering
\caption{Robustness and silent failure metrics under Gaussian corruption. The column `SF` denotes the Silent Failure Rate ($\downarrow$) and `F1` denotes the Attack-F1 ($\uparrow$).}
\label{tab:gaussian}
\resizebox{0.95\textwidth}{!}{%
\begin{tabular}{|l|l|cc|cc|c|c|c|c|cc|cc|}
\hline
\multirow{2}{*}{Dataset} & \multirow{2}{*}{Str.} & \multicolumn{2}{c|}{RES-DARE} & \multicolumn{2}{c|}{XGBoost} & \multicolumn{2}{c|}{Random Forest} & \multicolumn{2}{c|}{MLP} & \multicolumn{2}{c|}{LogReg} \\ \cline{3-12} 
 &  & F1 $\uparrow$ & SF $\downarrow$ & F1 $\uparrow$ & SF $\downarrow$ & F1 $\uparrow$ & SF $\downarrow$ & F1 $\uparrow$ & SF $\downarrow$ & F1 $\uparrow$ & SF $\downarrow$ \\ \hline
\multirow{2}{*}{CICIDS17} 
 & 0.05 & \textbf{0.8699} & 0.0050 & 0.3549 & 0.3391 & 0.2507 & \textbf{0.0001} & 0.5019 & 0.4989 & 0.5007 & 0.4996 \\
 & 0.10 & \textbf{0.7920} & 0.0091 & 0.3410 & 0.3434 & 0.2155 & \textbf{0.0003} & 0.5015 & 0.4996 & 0.5003 & 0.5001 \\ \hline
\multirow{2}{*}{UNSW15} 
 & 0.05 & \textbf{0.9720} & 0.0004 & 0.9122 & 0.0558 & 0.9267 & \textbf{0.0000} & 0.9663 & 0.0117 & 0.8986 & 0.0136 \\
 & 0.10 & \textbf{0.9547} & 0.0015 & 0.8928 & 0.0736 & 0.9083 & \textbf{0.0000} & 0.9490 & 0.0244 & 0.8661 & 0.0414 \\ \hline
\multirow{2}{*}{TON\_IoT} 
 & 0.05 & \textbf{0.9311} & 0.0058 & 0.6954 & 0.0669 & 0.8352 & \textbf{0.0001} & 0.8720 & 0.0515 & 0.8812 & 0.0325 \\
 & 0.10 & \textbf{0.9044} & 0.0106 & 0.5334 & 0.1144 & 0.7490 & \textbf{0.0001} & 0.7543 & 0.1361 & 0.8210 & 0.0532 \\ \hline
\end{tabular}%
}
\end{table}

The corresponding feature-scaling drift results are reported in Table~\ref{tab:scaling}. RES-DARE remains competitive on CICIDS2017 and UNSW-NB15, but the TON\_IoT subset shows a clear weakness at strength 0.40, where Random Forest and MLP retain much higher Attack-F1. This confirms that the current expert-routing and repair mechanisms are more effective for noisy-feature degradation than for global feature-rescaling shifts.

\begin{table}
\centering
\caption{Attack-F1 under feature-scaling drift ($\uparrow$).}
\label{tab:scaling}
\begin{tabular}{|l|l|c|c|c|c|c|}
\hline
Dataset & Strength & RES-DARE $\uparrow$ & XGBoost $\uparrow$ & RF $\uparrow$ & MLP $\uparrow$ & LogReg $\uparrow$ \\
\hline
\multirow{2}{*}{CICIDS2017}
& 0.20 & 0.9600 & \textbf{0.9719} & 0.9386 & 0.6691 & 0.6648 \\
& 0.40 & 0.7945 & 0.5029 & 0.6115 & \textbf{0.9543} & 0.7214 \\
\hline
\multirow{2}{*}{UNSW-NB15}
& 0.20 & \textbf{0.9505} & 0.9144 & 0.9325 & 0.9414 & 0.8915 \\
& 0.40 & \textbf{0.9449} & 0.8595 & 0.9385 & \textbf{0.9445} & 0.8782 \\
\hline
\multirow{2}{*}{TON\_IoT}
& 0.20 & 0.8795 & 0.9212 & \textbf{0.9437} & 0.9345 & 0.8997 \\
& 0.40 & 0.6386 & 0.6529 & \textbf{0.9175} & 0.9100 & 0.8999 \\
\hline
\end{tabular}
\end{table}

\subsection{Warning Signals and Interpretability}

For a deployed IDS, knowing that the model is about to fail is valuable. Table~\ref{tab:uncertainty} compares four uncertainty signals on their ability to separate correct predictions from incorrect ones, measured by warning AUROC. Monte Carlo Dropout (MC Dropout) estimates predictive uncertainty by performing multiple stochastic forward passes with dropout enabled at inference time and measuring variability across predictions \cite{gal2016dropout}. While principled, it requires several forward passes per prediction, making it computationally expensive. Deterministic confidence and entropy match or outperform MC Dropout across all three datasets, meaning the cheaper signals work just as well here.

The model is also more confident when it is right than when it is wrong: mean confidence is 0.9497 versus 0.6783 on CICIDS2017, 0.9182 versus 0.6035 on UNSW-NB15, and 0.9282 versus 0.6311 on TON\_IoT. Feature-occlusion analysis was used as a diagnostic during development, but it is not reported as a main interpretability result here because the processed feature-index mapping must be validated before feature-level claims are made.

\begin{table}
\centering
\caption{Error-warning AUROC. Higher is better ($\uparrow$). TON\_IoT rows use TON\_IoT\_SMALL\_5P; uncertainty evaluation sampled $n=20{,}000$.}
\label{tab:uncertainty}
\begin{tabular}{|l|c|c|c|c|}
\hline
Dataset    & $1-\mathrm{conf}$ $\uparrow$ & Entropy $\uparrow$ & MC variance $\uparrow$ & MC entropy $\uparrow$ \\
\hline
CICIDS2017 & 0.9760 & 0.9760 & 0.9573 & \textbf{0.9761} \\
UNSW-NB15  & \textbf{0.9595} & \textbf{0.9595} & 0.8580 & 0.9587 \\
TON\_IoT   & \textbf{0.9511} & \textbf{0.9511} & 0.8121 & 0.9504 \\
\hline
\end{tabular}
\end{table}

\FloatBarrier

\section{Discussion}

The central finding of this work is that clean accuracy is an incomplete measure of IDS quality. Static baselines such as XGBoost and Random Forest are highly competitive on clean tabular benchmarks and should be expected to win those comparisons. RES-DARE does not dispute this. What it adds is the layer beneath: graceful degradation under noise, warning signals before failure, expert adaptation as distributions shift, and rollback-safe repair when adaptation is needed. These properties are what determines whether a system remains trustworthy after deployment.

\subsection{Limitations}

The main limitation is scaling drift resistance. Under synthetic feature scaling at strength 0.40, RES-DARE drops to 0.6386 Attack-F1 on the TON\_IoT subset, while Random Forest and MLP remain above 0.90. The current expert-routing and repair mechanisms handle noisy-feature degradation well but are less effective against global feature-rescaling shifts. AEHM-v2 validation was observed on all three datasets, but the TON\_IoT repair-validation run used a stratified 5\% subset rather than the full TON\_IoT evaluation set. The current checkpoint format also does not fully restore all expert-object state, limiting sample-level routing entropy analysis at evaluation time. Baseline comparison is limited to classical static baselines and a reported-value QCL-IDS comparison, so broader same-protocol comparison against recent adaptive IDS methods is still needed. Finally, feature-occlusion diagnostics require validated feature-name mappings before they can support dataset-level interpretability claims.

\subsection{Future Scope}

Future work will extend repeated-seed analysis to the robustness and corruption experiments and, when computationally feasible, to repeated full TON\_IoT runs. Scaling-drift robustness should be improved through normalisation-aware routing or scaling-robust encoder training. Broader baseline comparison should include recent continual and adaptive IDS methods under a unified evaluation protocol. The implementation should also improve full expert-state checkpointing so that sample-level routing entropy and richer post-hoc analysis can be restored from saved models. Finally, validated feature-name mappings and richer explainability methods such as SHAP should be added before making detailed feature-level claims, and deployment should be evaluated on real network infrastructure outside benchmark conditions.

\section{Conclusion}

Intrusion detection systems are trusted to operate in conditions they were never trained on, against adversaries who are actively incentivised to find their blind spots. RES-DARE is built around the argument that the right response to this is not to chase perpetual autonomous operation, but to extend the reliable operational window as far as possible and make its limits visible. The results support this position: competitive clean detection, near-zero forgetting, substantially better robustness under feature corruption, and warning signals that reliably precede failure. AEHM-v2 ensures that adaptation is never a gamble -- every repair is validated before it is committed, and rolled back if it is not. Taken together, these properties describe a system that does not just perform well on a benchmark, but behaves honestly under pressure.

\section*{Acknowledgments}
The authors gratefully acknowledge the support of the Variable Energy Cyclotron Centre (VECC), the Department of Atomic Energy (DAE), Government of India, for providing the infrastructure and technical environment that supported this research. The authors also thank the staff of the VECC library for their assistance during the course of this study.

\section*{Declarations}

\subsubsection*{Consent to Publish}
All authors have read and approved the final manuscript and consent to its submission and publication.

\subsubsection*{Conflict of Interest}
The authors declare that they have no known competing financial interests or personal relationships that could have appeared to influence the work reported in this paper.

\bibliographystyle{IEEEtran}
\bibliography{references}

\end{document}